\documentclass[twocolumn,showpacs,preprintnumbers,pra,aps,10pt,superscriptaddress]{revtex4-1}

\usepackage{amsmath}
\usepackage{amsfonts}
\usepackage{amssymb}
\usepackage{graphicx}
\usepackage{color}
\usepackage{enumerate}

\newcommand{\ket}[1]{\left| #1 \right>} 

\begin{document}
\title{Efficient entanglement length measurements for photonic cluster state sources}

\author{Ilai Schwarz}
\affiliation{Racah Institute of Physics, The Hebrew University, Jerusalem 91904 Israel}

\author{Terry Rudolph}
\affiliation{Controlled Quantum Dynamics Theory Group, Imperial College London, London SW7 2AZ, United Kingdom}

\begin{abstract}
We present a procedure for confirming the generation of long strings of photons in an entangled (cluster) state that does not rely on complete state tomography and that works even at low collection efficiencies. The scheme has the added advantage of being passive - it does not require switching of optical elements to perform measurements in different bases for instance.

\end{abstract}
\maketitle

Recently, a scheme for preparing on-demand photonic cluster state strings (a ``cluster state machine gun'') was proposed \cite{lindner_proposal_2009} and adapted to various architectures \cite{nielsen_deterministic_2010,lin_efficient_2010,ballester_generation_2011,li_photonic_2011}. Such a device would ideally allow for the production of hundreds of entangled photons entangled in a 1D cluster state, a technology that would dramatically change the face of optical quantum information processing. However, it immediately brings to the fore the challenge of benchmarking and characterizing the output light: How does one verify that the emitted photons are actually entangled as expected, and how many photons are entangled before inevitable noise destroys the quantum correlations?

The simplest solution would be to perform a full tomography on $K$ photons to deduce the reduced density matrix $\rho_K$ of these $K$ photons. From $\rho_K$ one can try and calculate the entanglement between the photons using known entanglement measures. The problem is that given current experimental setups, $K$ must be quite small because of collection \& photodetection efficiencies which reduce the number of sequential photons that are measured.


For example, denoting by $p_d$ the probability of photon collection and detection, the probability of detecting $K$ photons in a row is $(p_d)^K$. Furthermore, there are $2^{2K}$ variables in the density matrix of $K$ photons. In order to do the full tomography, even given a good passive experimental setup, one will need at least $2^{2K}$ measurements of $K$ photons in a row
to get a measurement of every one of the variables (actually a lot more, but this will be a lower bound). Therefore, we get that the amount of data needed for one measurement of each variable is bounded below by
\begin{equation}\label{naiveTomography}
    N = \frac{2^{2K}}{p_d^K},
\end{equation}
with $N$ being the number of photons emitted. For one photon emitted per nanosecond with collection efficiencies of $10\%$, $50\%$ and $90\%$ and a measurement time of $10$ seconds, this yields $6,\;11,\;15$  as the best case number of entangled photons in the reconstructed cluster state.


\begin{figure}
\centering
\includegraphics[width=0.9\columnwidth]{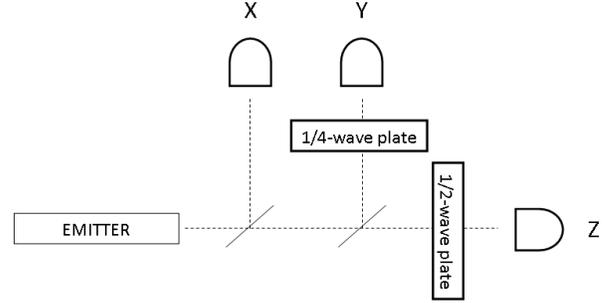}
\caption{\label{fig:ex_setup} The proposed experimental setup for measuring the entanglement length in a passive manner. The reflectivity of the beam-splitters can be adjusted so the probability of measuring the photon in each of the Pauli bases is controllable. The default detector polarization basis is assumed to be the horizontal and vertical polarizations.}
\end{figure}

In addition to the issues of collection efficiencies, short (nanosecond) time scales between photons are needed for the quantum dot example considered in Ref.~\cite{lindner_proposal_2009} to achieve minimal decoherence. As such allowing active optical elements that can be switched between photons in the setup will be very difficult.

In this paper we propose an experimental setup that allows one to perform the tomography \emph{passively}, and show a specific method in which the entanglement of a string of photonic cluster state can be measured directly for up to $6,\; 20,\; 80$ photons for the same emission/collection parameter values mentioned above.  More importantly, we will show that from the same measurement data it is possible to indirectly infer entanglement between several hundreds of photons, even with standard Pauli noise rates per photon at the several percent level.

The trick will be to devise a scheme that does \emph{not} require a large number of successful detections in a row.
Fig.~\ref{fig:ex_setup} shows our proposed experimental setup for the tomography. The emitted photons pass through two beam splitters, randomly ending up in one of three detectors. One beam goes into a detector measuring circular polarization (eigenvectors denoted as $\ket{L}$, $\ket{R}$), one in horizontal or vertical polarization ($\ket{H}$,$\ket{V}$), and the third in diagonal polarization ($\ket{D}$,$\ket{\bar{D}}$). These three polarization bases will be denoted as the $Z$ basis, $X$ basis and $Y$ basis  respectively. The translation from one polarization basis to another can be done by quarter-wave plates and half-wave plates, as depicted in Fig.~\ref{fig:ex_setup}.

The experiment now proceeds as follows: photons are emitted from the cluster state machine gun, which can be either a quantum dot in a microcavity \cite{lindner_proposal_2009} or any other similar setup \cite{nielsen_deterministic_2010,lin_efficient_2010,ballester_generation_2011,li_photonic_2011} which emits photons entangled in a linear cluster state. These photons then go one by one through the beam splitters, which act classically and just randomly direct each photon towards a detector. Denoting by $X$, $Y$ and $Z$ a photon measured in the corresponding basis, and by $\_\_$ a photon which was lost because of collection efficiency, a typical measurement result will look like a random string of `$X$',`$Y$' , `$Z$' and `$\_\_$' together with the measured eigenvalues:
\[
 \begin{array}{ccccccccc}
   X & \_\_ & Z & \_\_ & \_\_ & Z & X & Y & \_\_ \\
   +1 & \_\_  & -1  & \_\_ & \_\_  & -1 & +1 & +1 & \_\_
 \end{array}
 \]
In some of what follows measurements in only two bases are needed which can be achieved with one less beamsplitter.

In order to show that this passive measurement setup suffices to give us all the information needed for measuring the entanglement length we make the following assumptions: First, we presume the state of the emitted photons is translationally invariant; i.e. the state of photons $i,\ldots,i+l$ is identical to the state of photons $j,\ldots,j+l$, for every $i,\;j,\;l$. This amounts to assuming a continuous firing of the machine gun for long time periods is possible, but does not amount to assuming that the error mechanisms are slow over such periods. The second assumption is that the errors of the state are independent Pauli errors on one or perhaps two adjacent photons. In the proposed scheme for the pulsed optical cluster state~\cite{lindner_proposal_2009}, as well as in most other setups, both these assumptions hold - they are basically a property of the cluster state being created.

The localizable entanglement (LE) between two particles is a measure of the maximal possible bipartite entanglement that can be created between them by local operations on the other particles with which they are correlated \cite{popp_localizable_2005,verstraete_entanglement_2004}. In perfect linear cluster states the distance over which the LE  is non-zero is infinite.  For an imperfect linear cluster state output from the machine gun, our goal will be to determine the distance (number of in-between photons) over which the LE between photons is non-zero. We define this quantity as $\xi_E$. Therefore finding $\xi_E>0$ gives  an operational indication of whether two particles share useful entanglement.

Imagine now that measurement of the localizable entanglement $LE(k,k+l)$ between photon $k$ and photon $k+l$, which are $l$ photons apart, is attempted. Denoting by $\mathcal{M}$ a measurement on all the photons except photons $k, \; k+l$, (the measurement can include photons which are not really measured, but just traced out), it must hold that
\begin{equation}\label{LE}
    LE(k,k+l) \geq \sum_s p_sE(\rho^s_{k,k+l})\;,
\end{equation}
with $s$ running over the outcomes of the measurement $\mathcal{M}$, $p_s$ is the probability of said outcome and $E(\rho^s_{k,k+l})$ is the entanglement between photons $k, \; k+l$, which can be deduced from the reduced density matrix. Thus, this gives us a lower bound on the localizable entanglement.
We will be  interested in measurements that have the same entanglement for the 2 photons $k$, $k+l$ for all measurement outcomes, thus reducing Eq.~\eqref{LE} to $ LE(k,k+l) \geq E(\rho_{k,k+l})$. These measurement sequences do not change depending on previous measurement outcomes, and can therefore be measured passively (see Ref.~\cite{doherty_identifying_2009} for a more detailed discussion). Also, because the cluster state has translational invariance (its stabilzer group is translationally invariant), we can denote $\rho^l_{12} = \rho_{k,k+l}$, for all values of $l$, with $\rho^l_{12}$ now being the reduced density matrix for two photons distance $l$ apart. Eq.(~\ref{LE}) now becomes
\begin{equation}\label{LE2}
    LE(k,k+l) \geq E(\rho^l_{12}).
\end{equation}
In order to calculate the entanglement $E(\rho^l_{12})$, since the reduced density matrix will typically be mixed  (as we have traced over some of the other photons plus there is possibly noise), we will use the entanglement of formation \cite{bennett_mixed-state_1996, wootters_entanglement_1998} for which an explicit formula is available.

The question we now face is: what  measurement sequence $\mathcal{M}$ will maximize Eq.~\ref{LE2}, and give us the best lower bound on the LE? Since the experimental state will be a cluster state with additional errors, let us first explain how to get the best lower bound on the $LE$ for the perfect cluster state.

\begin{figure}
\centering
\includegraphics[width=0.9\columnwidth]{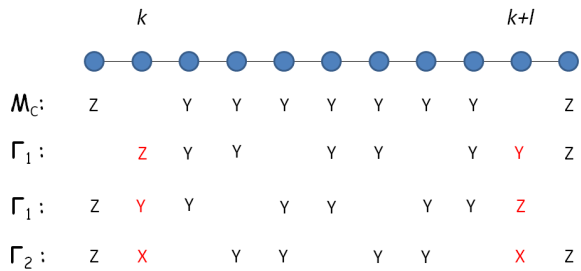}
\caption{\label{fig:MC_stabilizers} (color online) The figure depicts the measurement sequence $\mathcal{M}_C$ and the cluster state stabilizers which are responsible for the entanglement between the two unmeasured photons $k,\;k+l$ for $l=8$. The Pauli matrices depicted in red are the stabilizers of the photons $k,\;k+l$ after the measurement sequence $\mathcal{M}_C$.}
\end{figure}

For the ideal cluster state \cite{raussendorf_measurement-based_2003}, we  use the special properties it possesses from being a stabilizer state \cite{gottesman_stabilizer_1997}, with the stabilizers $K_i = Z_{i-1}X_iZ_{i+1}$. A set of Pauli measurements denoted $\mathcal{M}_C$ is shown in Fig.~\ref{fig:MC_stabilizers} that has the property it collapses qubits $k$ and $k+l$ into a maximally entangled state (similar to the measurement set introduced in Ref.~\cite{hein_entanglement_2006}). To see why, consider a pattern of products of stabilizers of the form $K_1 K_2 K_4 K_5 K_7 K_8...$. This yields a sequence of Pauli operators depicted as $\Gamma_1$ in Fig.~\ref{fig:MC_stabilizers}. If we shift this sequence one qubit to the left (recall the state is translationally invariant) we obtain  the second form of $\Gamma_1$ depicted. Consider also a  pattern of products of stabilizers of the form $K_{1}K_3K_4K_6K_7...$, which yields a pattern $\Gamma_2$, also shown in the figure. Both $\Gamma_1$ and $\Gamma_2$ must yield a value $+1$ when measured on the cluster state, since they are products of stabilizers, and both can be found as sub-sequences of $\mathcal{M}_C$.  From this one deduces that the resultant state on qubits $k$ and $k+l$ in the figure are stabilized by $\{YZ,ZY,XX\}$, which are the stabilizers of a maximally entangled state.



Even when the relevant state is not the perfect cluster state we will show the measurement sequence $\mathcal{M}_C$ gives a very good lower bound on $\xi_E$ \footnote{as long as the state still has a $Z\rightarrow -Z$ symmetry, as explained in Ref.~\cite{venuti_analytic_2005}.}. In this case, the entanglement length will be finite, but can still be large.

There are two parts to our proposal for measuring $\xi_E$. The first involves a direct determination of (a lower bound on) $E(\rho^l_{12})$ for any given $l$ by finding appropriate sequences of measurements in the data. The second idea is to use a more detailed analysis of how the expectation values of certain sequences in the data are changing with the length of the sequence to determine the rate of various errors on the cluster state. If the model fits (which can be verified) then very large values of  $\xi_E$ can then be inferred. We refer to these as the direct and indirect methods for inferring $\xi_E$ and begin by analysing the direct method.

Key to our proposal is that instead of looking for the full measurement sequence $\mathcal{M}_C$, as discussed above we can measure $\langle\Gamma_1\rangle$ and $\langle\Gamma_2\rangle$. That is, taking the very long string of measurements in random bases outputted from the experimental setup of Fig.~\ref{fig:ex_setup}, one can expect to sometimes find instances of the sequence $\Gamma_1$ or $\Gamma_2$ from which these expectations can be determined. Both of these sequences allow for certain qubit locations to remain undetermined. That is, such photons can be lost without detriment - we require less photons ``in a row'' to be collected. As is clear from Fig.~\ref{fig:MC_stabilizers}, $\left<\Gamma_1\right>$ gives us the values for $Tr\left[\rho^l_{12} (Z\otimes Y)\right]\equiv \mu_{ZY}$ and $Tr\left[\rho^l_{12} (Y\otimes Z)\right]\equiv \mu_{YZ}$, while $\left<\Gamma_2\right> = Tr\left[\rho^l_{12} (X\otimes X)\right]\equiv \mu_{XX}$.

With these values in hand we can find a lower bound for $E(\rho^l_{12})$. In fact
$E(\rho^l_{12}) \ge E(\tilde{\rho}^l_{12})$ where we define
\begin{equation}\label{rho_tag}
    \tilde{\rho}^l_{12} = \frac{1}{4}(I + \mu_{YZ} Y\otimes Z + \mu_{ZY} Z\otimes Y + \mu_{XX} X\otimes X).
\end{equation}
That is, although the actual state of the two qubits will (since it is imperfect cluster state) contain other terms in the density matrix, these can only increase the possible entanglement of the state. To see this note that an LOCC protocol (which cannot increase entanglement) in which the three operators $ZY,YZ$ and $XX$ are each randomly applied to the two qubits always brings $\rho^l_{12}$ to the form of $\tilde{\rho}^l_{12}$. Note also that not only are we not having to perform full tomography on the $l$ qubits to demonstrate their entanglement, we are not even performing full tomography on qubits $k,k+l$.

We now calculate the number of photons needed for a single instance of a useful measurement sequence of length $l$. Both $\Gamma_1,\; \Gamma_2$ have a preferred  measurement direction $Y$, in which the number of measured photons increases with the length of the sequence. Denote by $p_p$ the probability that the beam splitter passes the photon to the $Y$-direction detector and by $p_d$ the probability of photon detection. Since the probability of getting even a start of a string is small, we can assume that the mean number of photons needed for one instance of the string is $1/p_l$, with $p_l$ being the probability of getting a sequence of length $l$.  We get that
\begin{equation}\label{dataNeeded}
    p_l = p_d^{n_m}p_p^{n_p}\left(\frac{1-p_p}{a}\right)^{n_m-n_p},
\end{equation}
with $n_m$ being the total number of photons measured, $n_p$ the number of photons measured in the preferred direction ($Y$) and $a$ is the number of detectors in the experimental setup not in the preferred direction ($a=1$ for $\Gamma_1$, and $a=2$ for $\Gamma_2$).
Now, for each sequence, one can find the optimal value of $p_p = \frac{n_p}{n_m}$ by derivation.  Plugging this back into Eq.~\ref{dataNeeded} gives
\begin{eqnarray}\label{dataCompact}
  p^{\Gamma_1}_l &=& \left(\frac{3}{l+1}\right)^2 \left(p_d\frac{l+1}{l+4}\right)^\frac{2l+8}{3}, \nonumber \\
  p^{\Gamma_2}_l &=& \left(\frac{3}{l-2}\right)^4 \left(p_d\frac{l-2}{l+4}\right)^\frac{2l+8}{3}, \nonumber
\end{eqnarray}
for $l > 2$.

\begin{figure}
\centering
\includegraphics[width=0.9\columnwidth]{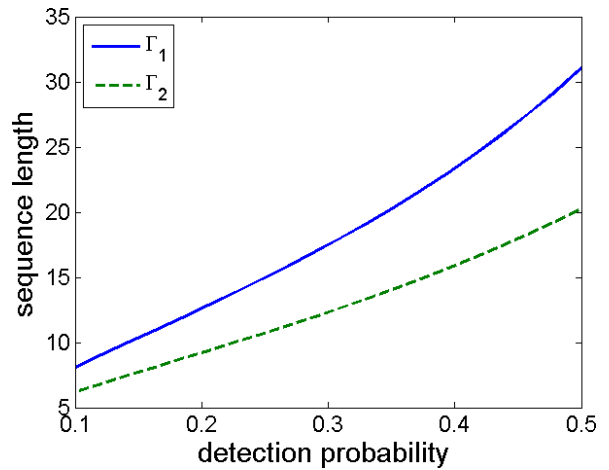}
\caption{\label{fig:datalength} The maximum instance length for each one of the sequences $\Gamma_1, \; \Gamma_2$ which will be measured by the direct method in the experimental setup (depicted in Fig.~\ref{fig:ex_setup}) as a function of the photodetection+collection probability. The rate of emitted photons is taken as 1 photon every nanosecond, and the measurement time is 10 seconds. }
\end{figure}

Therefore, since the number of photons needed for a sequence $\Gamma$ is the corresponding $1/p^\Gamma_l$, the amount of time needed for one instance of length $l$ is $\tau_{em} / p^\Gamma_l$, with $\tau_{em}$ being the the average time for the emission of one photon (and is proportional to $T_{cycle}$ for the pulsed cluster state scheme).
Fig.~\ref{fig:datalength} shows the maximum instance length for each one of the sequences $\Gamma_1, \; \Gamma_2$ which will be measured in the experimental setup (depicted in Fig.~\ref{fig:ex_setup}) in a reasonable time (10 seconds) as a function of the photodetection+collection probability. This assumes $\tau_{em} = 1\;ns$ for the quantum dot (or the emission rate for other experimental emitters). We see that for reasonable photodetection+collection probabilities, sequences of length greater than 15 can be measured directly.

The direct method above gives first hand information about the LE for the measured sequence lengths. We now describe an indirect method for extrapolating from the same random sequence of measurements a much higher lower bound for $\xi_E$.

To do this we need to consider the possible errors on the linear cluster state. The two scenarios which will be investigated are the case of single random Pauli errors, and the case of $Z_i\otimes Z_{i+1}$ errors, both of which are relevant for the photonic cluster state strings \cite{lindner_proposal_2009}. In fact, as was shown there, single qubit Pauli errors in the emitter transfer into being $Z_i\otimes Z_{i+1}$ errors on the output photons. We should therefore expect these pairwise Pauli errors to be the most relevant. Once a photon is fired out then there is no mechanism to generate single Pauli errors on it (there is no populated bath it couples to at optical frequencies). However, given the variety of systems being considered to build a machine gun it could well be that different schemes to quantum dots do produce a small single Pauli error rate on the photons, and so we consider this case as well.

\begin{figure}
\centering
\includegraphics[width=0.9\columnwidth]{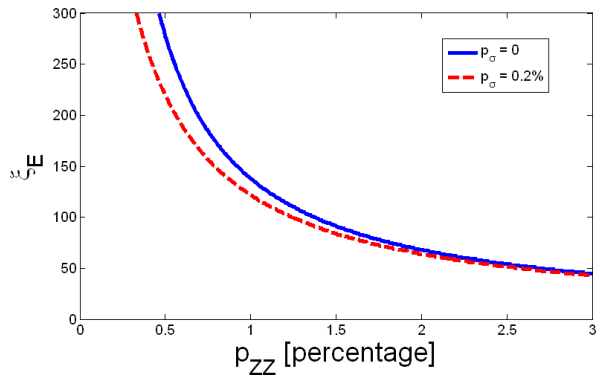}
\caption{\label{fig:entanglementlength} The localizable entanglement length $\xi_E$ as a function of the error probability $p_{ZZ}$, for two different values of Pauli error probability $p_\sigma = 0, \; p_\sigma=0.2\%$.}
\end{figure}

For single Pauli errors (i.e. each error acts as applying a Pauli operator on a single site), with the error probability $p_{\sigma}$, let us understand what value one should expect for a measurement sequence $\Gamma$ of length $l$. Assuming a uniform probability for the three different Pauli errors, for each of the measured photons two out of the three Pauli operators will cause $\Gamma$ to flip sign. This is true for any odd number of such errors. However, an even number of such errors on the measured photons will cancel out. Using the binomial theorem we get that
\begin{equation}\label{p_q}
    \left<\Gamma\right> = \left(1-\frac{4}{3}p_{\sigma}\right)^{n_m},
\end{equation}
with $n_m$ being the total number of photons measured in the sequence $\Gamma$. Therefore, in this case, the less photons needed to be measured for a sequence of length $l$, the higher the expectation value. Thus, all the expectation values for the different measurement sequences will decrease as $l$ increases, but at different rates corresponding to the number of measured photons for each measurement sequence.
Now, from the value for $\left<\Gamma\right>$ for different values of $l$ calculated from the experimental measurement string, it is possible to extract $p_{\sigma}$ (and see by the fit to the experimental data if this is the right model to use). Given $p_{\sigma}$ and the values of $\Gamma_1,\; \Gamma_2$, it is possible to extract the localizable entanglement for any length $l$, and find the value of $\xi_E$.

For the case of $Z\otimes Z$ errors, again assuming uncorrelated errors with a uniform error probability $p_{ZZ}$, we get an expectation value for each of the sequences that decreases exponentially with the length $l$:
\begin{eqnarray}\label{ZZ_errors}
  \left<\Gamma_1\right> &=& \left(1-2p_{ZZ}\right)^{\frac{2}{3}l}, \nonumber \\
  \left<\Gamma_2\right> &=& \left(1-2p_{ZZ}\right)^{\frac{2}{3}l}. \nonumber
\end{eqnarray}

As for the single Pauli error case, we can use sequences of varying length $l$ to infer the error rates and then lower bound the localizable entanglement length $\xi_E$. Fig.~\ref{fig:entanglementlength} depicts the bound we can obtain as a function of the error probability $p_{ZZ}$, for two different values of single qubit Pauli error probability $p_\sigma = 0, \; p_\sigma=0.2\%$. It is important to note that this indirect method yields a much tighter lower bound than that which would be achieved by bounding $\xi_E$ directly by measuring only the cluster state stabilizers $K_i = Z_{i-1}X_iZ_{i+1}$. We see that even for high error rates it will be possible to confirm the generation of long strings of entangled photons.


We conclude with a few observations. There are other possible measurement sequences to consider. For instance one can use stabilizers which correspond to measuring photons $k+1,k+2,\ldots,k+l-1$ in the $X$ basis, and the rest in the $Z$ basis \cite{skrvseth_phase_2009,hein_entanglement_2006}. These sequences actually give a tighter lower bound than $\Gamma_1,\; \Gamma_2$ in some cases. However, these sequences require measurements of more photons, so shorter sequences will be found in the measurement data at finite collection efficiencies. We make no claim as to the optimality of the measurement sequences we have chosen to illustrate our key ideas above.

Finally, one can also use the experimental measurement string outputted from the experimental setup for other purposes: reconstructing the pure quantum state with the highest fidelity with the experimental state \cite{cramer_efficient_2010}, or finding the fidelity of the experimental state with the ideal cluster state \cite{wunderlich_quantitative_2009, wunderlich_highly_2010}. Determining the efficiencies of these procedures as applied to the photon machine gun is an open problem.

\emph{
TR acknowledges the support of the UK Engineering and Physical Sciences Research Council and Toshiba Research Europe Limited.
}


\bibliographystyle{apsrev4-1}
\bibliography{entanglement_references}

\end{document}